\documentclass[12pt,epsf]{article}

\usepackage{latexsym}
\usepackage{epsfig}
\usepackage{amsfonts}

\begin{document}

\begin{center}
{\bfseries  \Large Casimir energy calculations within the
formalism of the noncompact lattice QED} \vskip 5mm

Oleg Pavlovsky$^{\dag}$ and Maxim  Ulybyshev$^{\ddag}$

\vskip 5mm

{\small {\it Institute for Theoretical Problems of Microphysics,
Moscow State University, Moscow, Russia }
\\
$\dag$ {\it E-mail: ovp@goa.bog.msu.ru } $\ddag$ {\it E-mail:
ulybyshev@goa.bog.msu.ru }}
\end{center}

\vskip 5mm

\begin{center}
\begin{minipage}{150mm}
\centerline{\bf Abstract} A new method based on the Monte-Carlo
calculation on the lattice is proposed to study the Casimir effect
in the noncompact lattice QED. We have studied the standard Casimir
problem with two parallel plane surfaces (mirrors) and oblique
boundary conditions on those as a test of our method. Physically,
this boundary conditions may appear in the problem of modelling of
the thin material films interaction and are generated by additional
Chern-Simons boundary term. This approach for the boundary condition
generation is very suitable for the lattice formulation of the
Casimir problem due to gauge invariance.
\end{minipage}
\end{center}

\vskip 3mm


\vskip 3mm


\vskip 10mm

\section{Introduction and general motivation}
During the last few years Casimir effect has attracted much
attention due to the great experimental and theoretical progress in
studying of this phenomenon. This macroscopic quantum effect  plays
crucial role in nanophysics, micro-mechanics, quantum optics,
condensed matter physics, material science, and also it is very
important for different models of boundary states in hadron physics,
heavy-ion collisions, and cosmology.

Nowadays, many theoretical methods for calculation of Casimir effect
are proposed. Unfortunately, most of these methods are based on the
fixed boundary conditions or external potentials. Such a
simplification can lead to problems in gauge invariance,
renormalizability and locality. Moreover, there are some problems
with different thermal corrections to Casimir force: different
methods of calculations of the Casimir effect predict different
corrections. Another part of the problem is calculation of Casimir
force for complicated forms of boundary surfaces. Typically, various
approximate methods (like the proximity force approximation method
\cite{bordag-mastepanenko}, \cite{Deryagin}) are used in the case of
curved surfaces, but it is still unclear now whether those
approximations are correct and for what tasks they can be applied.
And finally, analytic methods for Casimir effect calculation are
very complicated and strongly dependent on the shape of surfaces.
Practically, Casimir effect was studied analytically only for cases
of plane, spherical and cylindrical surface forms. More complicated
tasks have not been studied very well by this moment but such cases
typically appear in experiments.

Based on arguments discussed above, it seems a very important task
now to create a general method for calculation of Casimir effects
which would work well for different shapes of boundary surfaces,
at non-zero temperature and density and under other external
factors. It means that such a method should be formulated very
generally for working in different coupling regimes and different
external conditions. And we think that direct lattice calculations
in quantum field theory can meet all these requirements.

In our paper we consider the simplest Casimir problem with two
parallel planes as a test of our method. The existence of the
analytical answer for this problem is the additional motivation for
such a choice of bound surfaces.  This analytical  answer assists us
in fitting procedure for our numerical results.

 A crucial obstacle on the way of the realization of any Casimir problem on
the lattice is the following. If the Casimir energy is a reaction of
the vacuum on the presence of the boundary, what is "the boundary"
in terms of lattice formalism? In other words, what is an observable
quantity corresponding to such boundary? In fact, the answer is
non-trivial and we devote the first part of the paper to this
question. In the second part, we discuss the lattice algorithms and
numerical results. And the last part is a conclusion.

\section{Chern-Simons boundary conditions and Casimir effect}

Casimir effect is a reaction of the vacuum on boundary condition.
The spectrum of vacuum fluctuations depends on the boundary
conditions. Changing of the boundary conditions leads to changing of
the spectrum of vacuum fluctuations and so to generating of the
corresponding Casimir force on the boundary. In the standard quantum
field theory formalism, such changing of spectrum of vacuum
fluctuations can be described, for example, by means of Green
function method \cite{bordag-mastepanenko}. This approach is a very
powerful tool for studying many essential Casimir tasks
\cite{bordag-mastepanenko}. Unfortunately, the application of this
analytical method to the case of more complicated shape of the
boundary surfaces is not so easy due to calculation difficulties.
Our aim is the creation of the numerical method for the Casimir
effect calculation directly from the quantum field theory action.
The lattice formalism looks very attractive for this role but
manifestly we can not base in our approach on the separation of
vacuum modes corresponding with boundary from the full spectrum of
vacuum fluctuation. In lattice formalism we work in Euclidian space
and deal with full spectrum of vacuum fluctuations and can not
easily snatch out vacuum fluctuations corresponding to some boundary
conditions. We need some very delicate approach for separation of
vacuum fluctuation modes that preserve gauge invariance of our
lattice formalism. Fortunately, such an approach to Casimir problem
was proposed recently \cite{bordag-vasilevich, vasilevich}.

This approach is based on a very elegant idea coming from some
unique properties of the Chern-Simons action in three dimensions
\cite{bordag-vasilevich, vasilevich}. Let us consider
electro-magnetic fields in 3+1 dimensions with the Maxwell action
and additional Chern-Simons action given on 3-dimensional integral
on the boundary surface $S$:
\begin{equation}
S=-\frac{1}{4} \int d^4x\  F_{\mu\nu}F^{\mu\nu} -
 \frac{\lambda}{2} \oint d^3s \,
    \varepsilon^{\sigma \mu \nu \rho}
n_{\sigma } A_{\mu} ( x ) F_{\nu \rho} ( x )  , \label{action}
\end{equation}
where $\varepsilon^{\sigma \mu \nu \rho}$ is the Levi-Civita tensor
and $n_{\sigma }$ is  the normal vector to the boundary surface $S$,
$\lambda$ is a real parameter.

Let us consider now the simplest form of boundary surface $S$,
namely two parallel infinite planes placed at the distance $R$ from
each other. The Chern-Simons formulation of this canonical Casimir
problem was studied analytically in series of works \cite{markov1,
markov2}. We will use this analytical answer for the fitting of our
numerical data.

In the case of plane form of the boundary surface $S$ the
Chern-Simons action in (\ref{action}) has the following form:
\begin{equation}
 S_{CS}=\frac{\lambda}{2}\int(\delta(x_3)-\delta(x_3-{R}))
 \varepsilon^{3 \mu\nu \rho} A_{\mu}(x) F_{\nu
\rho}(x)d^4x. \label{actionCS}
\end{equation}
where, in our formulation of this Casimir problem, normal vectors to
the planes are turned in the opposite directions. This choice of the
normal vector orientation corresponds to our renormalization
procedure based on the connection between open and closed Casimir
problems.

If the parameter $\lambda$ is small, electro-magnetic fields
obviously don't feel any boundary and are free. What happens if the
parameter $\lambda$ becomes large and tends to infinity and  fields
dynamics on the boundary surface $S$ is determined by Chern-Simons
action? Let us consider the equation of motion obtained from the
action (\ref{action}):
\begin{eqnarray}
\Box A^\mu + \lambda (\delta(x_3)-\delta(x_3-{R}) )\varepsilon^{ 3
\sigma  \nu \rho} A_\sigma
\partial_\nu A_\rho  =0 . \label{equa_mot}
\end{eqnarray}
At $\lambda \rightarrow \infty$, it is easy to obtain from
(\ref{equa_mot}) a corresponding boundary conditions on the surface
$S$:
\begin{eqnarray}
&&E_\parallel \vert_{S}=0 ,\qquad H_n \vert_{S}=0 ,\label{condition}
\end{eqnarray}
where $H_n$ and $E_\parallel$ are normal and longitudinal components
of magnetic and electric fields correspondingly. These conditions
mean the nulling of the energy flux of the electromagnetic field
through the surface.
\begin{figure}[t]
\begin{center}
 \epsfbox{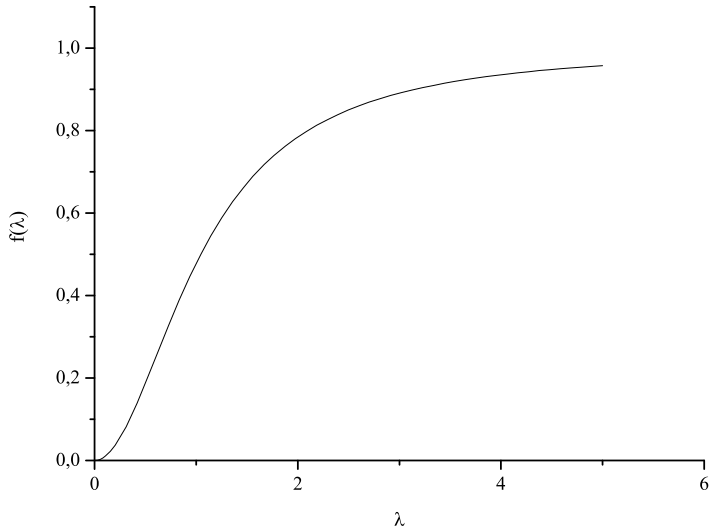} \end{center} \caption{ $f(\lambda)$. }
\end{figure}

This phenomenon corresponds to the well-known property of the
Chern-Simons theory in 3-dimension, namely to topological vortex
(strings) generation \cite{vortex, vortex2}. As it was shown by
Witten \cite{witten}, Chern-Simons theory, which is a quantum field
theory in three dimensions, is exactly solvable by using
nonperturbative methods and the topological vortices play a leading
role in this process. It was shown that the partition function of
this theory depends on the topology of the surface $S$ and gauge
group only, and tends to the zero if $\lambda$ tends to infinity.
There are no propagation modes in this case and if the boundary
surface $S$ is closed, the gauge fields inside and outside this
surface become separated from each other by the topological vortex
fields on the surface. In the case of the finite value of $\lambda$,
we have a non-trivial interaction between topological vortices and
electro-magnetic waves \cite{vortex2}. For our practical purposes it
is enough to know that the parameter $\lambda$ plays the role of the
effective regulator of the penetrability of our boundary surface for
electro-magnetic wave.

The analytical answer of Casimir energy per unit area for two planes
is given by Green function method is the following \cite{markov1,
markov2}:
\begin{eqnarray}
 E_{cas}=-\frac{\pi^2}{720 R^3} f(\lambda), \label{fullCas}
\end{eqnarray}
where function $f(\lambda): \lim_{\lambda \rightarrow \infty}
f(\lambda) = 1$ (Fig. 1) can be written as:
\begin{eqnarray}
 f(\lambda)=\frac{90}{\pi^4}\mbox{Li}_4\left(\frac{\lambda^2}{\lambda^2+1}\right). \nonumber
\end{eqnarray}
here the polylogarithm function $\mbox{Li}_4(x)$  is defined as
$$
\mbox{Li}_4(x)=\sum_{k=1}^\infty \frac{x^k}{k^4}=-\frac{1}{2}\int_0
^\infty k^2 \ln(1-xe^{-k})dk.
$$

In our paper we will study the Casimir energy per unit area behavior
at small $\lambda$, which is the following:
\begin{eqnarray}
 E_{cas}=-\frac{\lambda^2}{8 \pi^2 R^3}+ O(\lambda^4)
 \label{assimpt}
\end{eqnarray}

\section{Wilson "bag" and numerical lattice simulation of Maxwell-Chern-Simons theory}
\subsection{General statements}
In our paper we use the four-dimensional hyper-cubical lattice in
Euclidian space-time and the simplest form of the action for the
noncompact lattice QED:

$$
S=\frac{\beta}{2} \sum_x \sum_{\mu < \nu} \theta^2_{p,\mu\nu}(x),
$$
where the link and plaquette variables are defined as:
$$
\theta_{l,\mu}(x)=e\, a\, A_{\mu},
$$
$$\theta_{p,\mu \nu}(x)=\triangle_\mu \theta_{l,\nu}(x)-\triangle_\nu \theta_{l,\mu}(x),$$
$$\triangle_\mu \theta_{l, \nu}(x)=\theta_{l,\nu}(x+\hat{\mu}) - \theta_{l,\nu}(x).$$
Here $a$ is a lattice step and the parameter $\beta=1/e^2$. Physical
quantities are calculated in the lattice formalism by means of field
configuration averaging, where the field configurations (the set of
all link variables) are generated with the statistical weight
$e^{-S}$.

 We have clarified in previous sections that additional
Chern-Simons action describes Casimir effect. In order to find a
lattice description of Casimir interaction between boundary surfaces
let us consider Wilson loop, which describes the interaction of
charged particles. Wilson loop can be written in QED as
\begin{equation} W_C=e^{ig\oint \limits_C A_{\mu}dx_{\mu}}=e^{i\int J_{\mu}A_{\mu}dx^4}. \label{Wilsonloop} \end{equation}
The exponent in (\ref{Wilsonloop}) is the additional term to the
action. This term describes the interaction of the field $A_\mu$
with the current $J_{\mu}(x)=g\oint \limits_C
\delta(x-\xi)d\xi_{\mu}$ of charged particle. Configuration
averaging of Wilson loop $\langle W(R,T)\rangle$ (where R and T
are dimensions of the loop) converges in Euclidian time in the
limit $T\rightarrow\infty$ to:
$$\langle W(R,T)\rangle \rightarrow C e^{-V(R)T},$$ where $V(R)$ is the energy of interaction
between charged particles. The same method can be used for
calculation of Casimir energy by means of Chern-Simons action.

\begin{figure}[h]
 \begin{center} \epsfysize=60mm \epsfbox{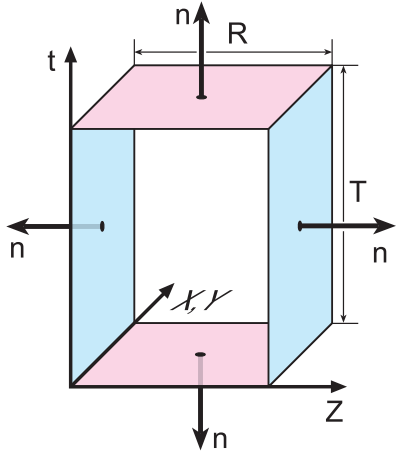}  \end{center}\caption{ Wilson bag for two plane surfaces. }
\end{figure}

Analogously to the description of charged particles interaction by
1D integral along Wilson loop, we will describe the Casimir
interaction of surfaces by corresponding 3D integral. The first
problem is that for stationary objects the action (\ref {actionCS})
is an integral from $t=-\infty$ to $t=\infty$, so by analogy to
Wilson loop, we should enclose the surface of the integration in
t-direction. The integration surface for two planes is shown in Fig.
2. This closing procedure can be performed both for plane surfaces
and for any curved surface in 3-dimensional space. As the result of
this procedure, so-called Wilson bag \cite{Seiberg,Luscher} can be
obtained. It can be written as
$$e^{i\lambda\oint\limits_{\Sigma}\varepsilon_{\mu\nu\rho\sigma}A_{\nu}F_{\rho\sigma}dS_{\mu}},$$
where $\Sigma$ is closed 3-dimensional surface in 4-dimensional
space-time. The final conclusion is that Wilson bag is (by analogy
to Wilson loop) observable quantity which gives us Casimir energy of
the objects, defined by the surface of integration. For two planes
we will calculate the following object:
\begin{equation}
 W_{Bag}(R,T)=e^{i \lambda S(R,T)},
 \label{refexp}
\end{equation}
 where $$S(R,T)= \int\limits_{0}^{T} dt
\int\!\!\int\!\!\int dx dy dz (\delta(z-R)-\delta(z))
\varepsilon_{3\nu\rho\sigma} A_{\nu}F_{\rho\sigma}+ $$
$$ +
\int\limits_{0}^{R}dz \int\!\!\int\!\!\int dx dy dt
(\delta(t-T)-\delta(t)) \varepsilon_{4\nu\rho\sigma}
A_{\nu}F_{\rho\sigma}.$$ And in the limit $T\rightarrow \infty$:
\begin{equation}
\langle W_{Bag}(R,T)\rangle \rightarrow C e^{-E_{cas}(R)T}.
\label{limbag}
\end{equation}

The second problem is to rewrite the Wilson bag in terms of
lattice objects (links and plaquettes). The product
$$A_{\nu}F_{\rho\sigma}$$
can be exactly constructed only in non-compact QED, because just
in the non-compact QED the link variable $\theta_{l, \nu}$ is a
lattice analogue of $A_{\nu}$ and the plaquette variable
$\theta_{p,\mu\nu}$ is a lattice analogue of $F_{\mu\nu}$.

There are two requirements for lattice representation of the
Wilson bag:

1) The whole integral for Wilson bag should be the gauge invariant
quantity.

2) "Locality". It means that in lattice representation of the
product $A_{\nu}F_{\rho\sigma}$, $A_{\nu}$ and $F_{\rho\sigma}$
should be given in the same point $x$. This requirement is
non-trivial, because $\theta_{p,\rho\sigma}$ gives the value of
$F_{\rho\sigma}$ in the center of the plaquette, but
$\theta_{\nu}$ gives the value of $A_{\nu}$ in the center of the
link. These are different points.

\subsection{Gauge invariance}
A basic element for building of the CS lattice action will be a
"corner" (Fig. 3). It is a product of the link variable and one of
the neighboring plaquette variables. This product is a lattice
analogue of the product $A_{\nu}F_{\rho\sigma}$.

\begin{figure}[t]
 \begin{center} \epsfysize=30mm \epsfbox{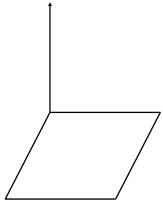}
 \end{center}
 \caption{ "Corner": the basic element for the building of CS lattice action.}
\end{figure}

First we will show how to construct the gauge invariant quantity
from these basic elements. Let's consider one CS plane closed over
the lattice period in all directions ($V_3$). The "naive"
discretization of this quantity in Euclidean space-time is
follows:
\begin{equation}
 S_{CS}=\sum_{x \in
V_3} \varepsilon_{\mu \nu \rho \sigma} n_{\mu} \theta_{l,\nu}(x)
\theta_{p,\rho \sigma}(x+\widehat{{dx}}_{\rho \sigma}),
\label{naiveCS}
\end{equation}
where the vector $\widehat{{dx}}_{\rho \sigma}$ appoints to the
plaquette that attached to the link $\theta_{l,\nu}(x)$:
$$
\widehat{{dx}}_{\rho \sigma}= c_{\rho \sigma} \hat{\nu} - a_{\rho
\sigma} \hat{\rho} - b_{\rho \sigma} \hat{\sigma},
$$
$$
a_{\rho \sigma}, b_{\rho \sigma} , c_{\rho \sigma}=0, 1.
$$

There are 8 variants how to join a link with one of the
neighboring plaquettes, two of them one can find in Fig. 4.
\begin{figure}[t]
 \begin{center} \epsfysize=60mm \epsfbox{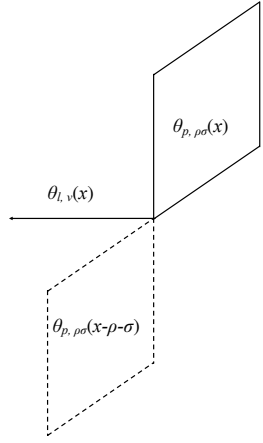}
 \end{center}
 \caption{ Two types of "corners".}
\end{figure}

\begin{figure}[b]
 \begin{center} \epsfysize=50mm \epsfbox{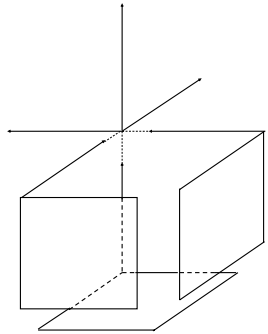}
 \end{center}
 \caption{ Combination of "corners" with different orientations that leads to the gauge invariant lattice Chern-Simons action.}
\end{figure}

Now let us consider gauge transformation of the "naive" lattice CS
action (\ref{naiveCS}).The gauge transformation of the link
variable is follows:
$$
\theta_{l,\nu} \rightarrow \theta_{l,\nu} + \alpha(x+\hat{\nu}) -
\alpha(x),
$$
and the action (\ref{naiveCS}) transforms into the following
expression due to the closing over the lattice period:
\begin{equation}
 S_{CS} \rightarrow \sum_{x \in
V_3} \varepsilon_{\mu \nu \rho \sigma} n_\mu \theta_{l,\nu}(x)
\theta_{p,\rho \sigma}(x+\widehat{dx}_{\rho \sigma}) + \sum_{x \in
V_3} \varepsilon_{\mu \nu \rho \sigma} n_\mu \alpha(x)
(\theta_{p,\rho \sigma}(x+\widehat{{dx}}_{\rho
\sigma}-\hat{\nu})-\theta_{p,\rho \sigma}(x+\widehat{dx}_{\rho
\sigma})). \label{transformCS}
\end{equation}

The second term in (\ref{transformCS}) appears due to the gauge
transformation. To formulate the lattice CS action in gauge
invariant form, this second term must be equal to zero. It is
possible to show that this term is equal to zero if plaquettes in
it are the sides of one 3d cube for all lattice sites $x$ and
normal vectors $n_\mu$. We should choose a variant to join a link
and a plaquette which has the property given above. One example of
such a join is shown in Fig. 5.

This combination of "corners" is organized as follows: for every
3d cube the "corners" of different orientations that lie inside it
(it means that "corners"  are built from edges of this cube) have
theirs link free ends being at one lattice site. This structure
can be obtained for example by the following choice of the vector
$\widehat{{dx}}_{\rho \sigma}$:
$$\widehat{{dx}}_{\rho \sigma}=  - \hat{\rho} -  \hat{\sigma},
$$
This choice provides the nulling of the expression:
$$\varepsilon_{\mu \nu \rho \sigma} (\theta_{p,\rho
\sigma}(x+\widehat{{dx}}_{\rho \sigma}-\hat{\nu})-\theta_{p,\rho
\sigma}(x+\widehat{dx}_{\rho \sigma}))=0, \forall x, \mu.
$$

Thus the Chern-Simons action obtained from such a system of
"corners" is the gauge invariant.

One can perform the similar consideration  for any closed
3-dimensional surface. An example of such a closing for the
"corner" with some orientation one can find in Fig. 6. Every
"corner" should be a part of such a chain and every chain should
be closed.
\begin{figure}[t]
 \begin{center} \epsfysize=100mm \epsfbox{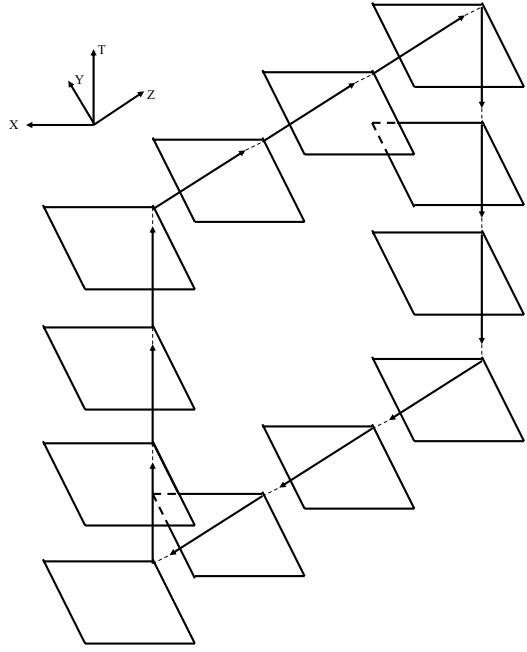}
 \end{center}
 \caption{ Closing for the "corner" of one orientation.}
\end{figure}
\subsection{Locality}
The next point of our consideration is the "locality". As it is
well known, the plaquette variable $\theta_{p,\rho \sigma}$ gives
us the value of $F_{\rho \sigma}$ in the center point of the
plaquette, while the link variable $\theta_{l,\mu}$ gives us the
$A_\mu$ in the middle point of the link. On the other hand, the
averages
$$
1/4
(\theta_{l,\nu}(x)+\theta_{l,\nu}(x+\hat{\rho})+\theta_{l,\nu}(x+\hat{\sigma})+\theta_{l,\nu}(x+\hat{\rho}+\hat{\sigma}))
$$
and
$$
1/2 (\theta_{p,\rho \sigma}(x)+\theta_{p,\rho \sigma}(x+\hat{\nu}))
$$
give us the values of $A_\nu$ and $F_{\rho \sigma}$ in the center
of the 3-dimensional cube determined by the 3 vectors
$\hat{\rho}$, $\hat{\nu}$ and $\hat{\sigma}$ based on one site
$x$.

It's obvious that the product of these two averages is the same as
the sum of all 8 variants of "corners", in other words we should
use all possible kinds to join of a link and a plaquette inside
one 3d cube. These 8 variants are shown on Fig. 7 for one possible
orientation of a "corner". For another two possible orientations
one has a similar set of variants.

\begin{figure}[t]
 \begin{center} \epsfysize=120mm \epsfbox{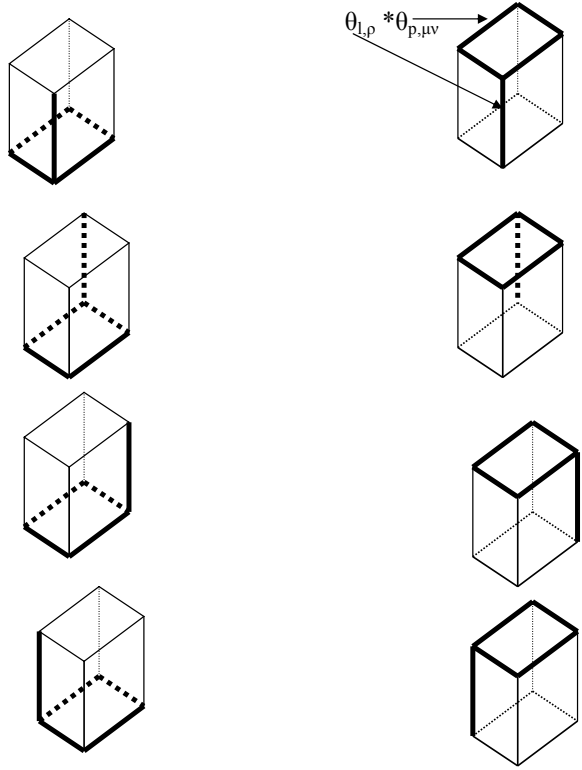}
 \end{center}\caption{ The structure of integral for "Wilson bag".
 To obtain the gauge invariant quantity we sum inside its 3-dimensional
 volume over all the products of plaquette and neighboring link. All these variants
 for one possible orientation of link and plaquette inside one
 3-dimensional cube are shown here.}
\end{figure}

Let us consider now all 3 orientation of "corners". For each of
them there is the set of 8 variants of possible join a link with a
plaquette. On other hand one can say that we have 8 systems of
"corners" (3 "corners" in one system). Each of these systems is
similar to the system of "corners" shown in the Fig. 5 (links'
free ends of all 3 "corners" are collected in one vertex of a
cube) and forms the gauge invariant quantity after summation over
the closed 3-dimensional surface.

Finally, our lattice representation of CS action is the sum over all
possible variants of the "corners" for three different orientations
inside the close 3-dimensional surface:
\begin{equation}
 S_{CS} = \frac{1}{8}\beta \sum_{x \in
V_3} \varepsilon_{\mu \nu \rho \sigma}
n_\mu(x)(\theta_{l,\nu}(x)+\theta_{l,\nu}(x+\hat{\rho})+\theta_{l,\nu}(x+\hat{\sigma})+\theta_{l,\nu}(x+\hat{\rho}+\hat{\sigma}))
(\theta_{p,\rho \sigma}(x)+\theta_{p,\rho \sigma}(x+\hat{\nu})) .
\label{latticeCS}
\end{equation}

The factor $1/8$ in (\ref{latticeCS}) comes from the averaging
over the 8 variants of the link and plaquette joining. The
expression $\theta_{l,\nu} \theta_{p,\rho \sigma}$ gives us $a^3
e^2 A_\nu F_{\rho \sigma}$ in continuum limit and by means of the
factor $\beta=1/e^2$ one can eliminate $e^2$ from the action.

\subsection{Deformed lattice}
To improve the accuracy we have used the lattice deformed in one
direction. This formalism is quite similar to one that was used in
article \cite{deformed} but instead of t-direction we deform one
of spatial directions (for definiteness z-direction). Briefly this
formalism can be described as follows:

All link variables are connected with the field $A_\mu$ by the
standard way:
$$\theta_{l,\mu}=e\, a\, A_{\mu}, \mu=1,2,4
$$
except links in z-direction:
$$\theta_{l,3}=\alpha e\, a\, A_{3}.
$$
Here $a$ is the lattice step in all directions except z and
$\alpha a$ is the lattice step in z-direction. The action of this
theory can be written in the following way:
$$S=\frac{\beta}{2} \Big\{ \alpha \sum_x \sum_{\mu < \nu; \mu, \nu \neq 3} \theta^2_{p,\mu\nu}(x) +\frac{1}{\alpha} \sum_x \sum_{\nu} \theta^2_{p,3 \nu}(x) \Big\}.
$$
The additional Chern-Simons action is not changed on the a such
deformed lattice. If $n_\mu$ (normal 4d vector to the 3d surface
of "Wilson bag") is oriented along the z-direction, deformed links
don't participate in "corners" and nothing changes. If $n_\mu$ is
oriented along one of non-deformed axes, one deformed link always
participates in link or plaquette variable of every "corner". For
example:
$$\theta_{l,3} \theta_{p,12} =e^2 \alpha a^3 A_3 F_{12},
$$
$$\theta_{l,2} \theta_{p,13} =e^2 \alpha a^3 A_2 F_{13}.
$$
But we don't need to divide by $\alpha$, because this automatic
presence of $\alpha$ in this expression simply corresponds to
change of elementary lattice 3-dimensional volume in this part of
"Wilson bag". Before deformation it was lattice 3d cube with
volume $a^3$. After deformation it's rectangular parallelepiped.
Its edges have lengths $a$, $a$ and $\alpha a$, so its volume is
$\alpha a^3$.

\section{Results of numerical calculations}
In our numerical simulations we calculate the Wilson bag quantity
(\ref{refexp}) with the lattice representation of $S_{CS}$ given
by the formula (\ref{latticeCS}). The Wilson bag surface is closed
over the lattice period in x- and y-directions. $E_{cas}$ is
extracted from the limit (\ref{limbag}).

First we study the dependence of $E_{cas}$ on $R$ (distance
between plates). In the Fig. 8 one can see $E_{cas} (R)$
dependence calculated on the lattice with dimensions 12 in x and
y-directions, 24 in z-direction (in order to decrease correlations
over the lattice period in z-direction) and 72 in t-direction (in
order to eliminate temperature effects). We did't use deformation
in this calculation. Parameter $\lambda$ is equal to $0.02$ here.

For this calculation we use $\beta=4$, but indeed in our Casimir
calculations nothing depends on $\beta$. Casimir effect without
radiational corrections doesn't depend on the electron charge in
the continuous theory. On the lattice the independence appears due
to the following reasons: 1) there is no phase transition in the
non-compact lattice QED and $\beta$ here only plays the role of
the scale parameter for a numerical value of link variables; 2) we
eliminate this dependence from $S_{CS}$ due to the multiplication
by $\beta$ in the final expression (\ref{latticeCS}). The absence
of the dependence was also tested numerically. So $\beta$ can be
chosen in rather a large interval according to calculational
convenience.

\begin{figure}[b]
\begin{center}
 \epsfbox{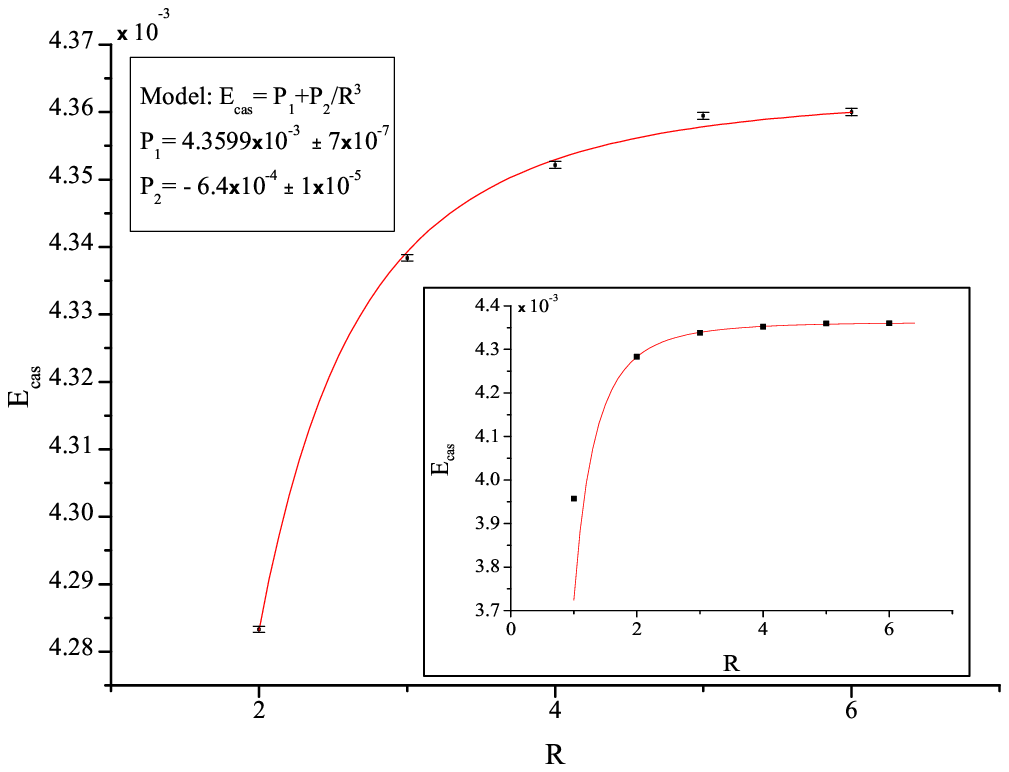} \end{center} \caption{$E_{cas}(R)$ at $\beta=4.0$ and $\lambda=0.02$.}
\end{figure}

It's well known in the continuous theory \cite{bordag-mastepanenko}
that $E_{cas}$ should depend on $R$ as $C/R^3$. So we'll try to fit
the point sequence obtained from the numerical simulation by the
function
\begin{equation}
E_{cas}=P_1+\frac{P_2}{R^3}. \label{model}
\end{equation}
 Our aim is, of course, the coefficient
$P_2$. $P_1$ is an analog of something like self energy of charged
particles. Such self energy can also be obtained in Wilson loop
calculation. The insertion in the Figure 8 presents the full
dependence from $R=1$ till $R=6$. The main graphic is the
increased segment from $R=2$ till $R=6$. The red line is the
fitting that was done for points $R=2...6$ on the model function
(\ref{model}). One can see that numerical points at lattice
distances $R=2...6$ lie on the fitting rather well. The point
$R=1$ differs from the normal $C/R^3$ dependence due to lattice
artefacts at small distances.

The values of coefficients $P_1$ and $P_2$ are given in the Figure
8. Now let us consider the physical meaning of the coefficient
$P_2$. A full Casimir energy of the interaction between two planes
have been obtained in these calculations. This energy is expressed
in units $a^{-1}$, where $a$ is the lattice step. If we take into
account that the area of the plane is equal to $(aN)^{2}$, where
$N=12$ is the size of the lattice in x and y-directions, we can
write the physical (dimensional) value of the Casimir energy
density:
$$E_{cas. phys}=\frac {1}{a} \frac {P_2}{R^3} \frac {1}{(aN)^2}=
\frac {P_2 N^{-2}} {(Ra)^3}.$$
After comparison between this
formula and (\ref{fullCas}) we can conclude that ${P_2 N^{-2}}$
should be equal to coefficient at $R^{-3}$ in (\ref{fullCas}) or
in (\ref{assimpt}) ($\lambda$ is small in our calculations by
reasons discussed later). And these values are rather close. $P_2
N^{-2} =-4.44 \times 10^{-6}\pm 7 \times 10^{-8}$ and the
theoretical value of this quantity for $\lambda=0.02$ is $-5.066
\times 10^{-6}$.

In order to improve the accuracy of calculation and obtain more
points for the fitting procedure, we consider the lattice deformed
in z-direction. As the planes are separated in z-direction, we'll
be able to study $E_{cas}(R)$ dependence more carefully. In the
Fig.9 one can see $E_{cas}(R)$ obtained on the lattice with
following parameters:
\begin{itemize}
\item 12 normal steps (the length of each step is $a$) in x and y
direction.
\item 72 normal steps in t-direction.
\item 120 small steps (the length of each one is $\alpha a$, $\alpha=0.2$) in
z-direction.
\end{itemize}

 So the whole volume and dimensions of this
lattice are the same that the previous one. The only difference is
smaller step in z-direction.

\begin{figure}[t]
\begin{center}
 \epsfbox{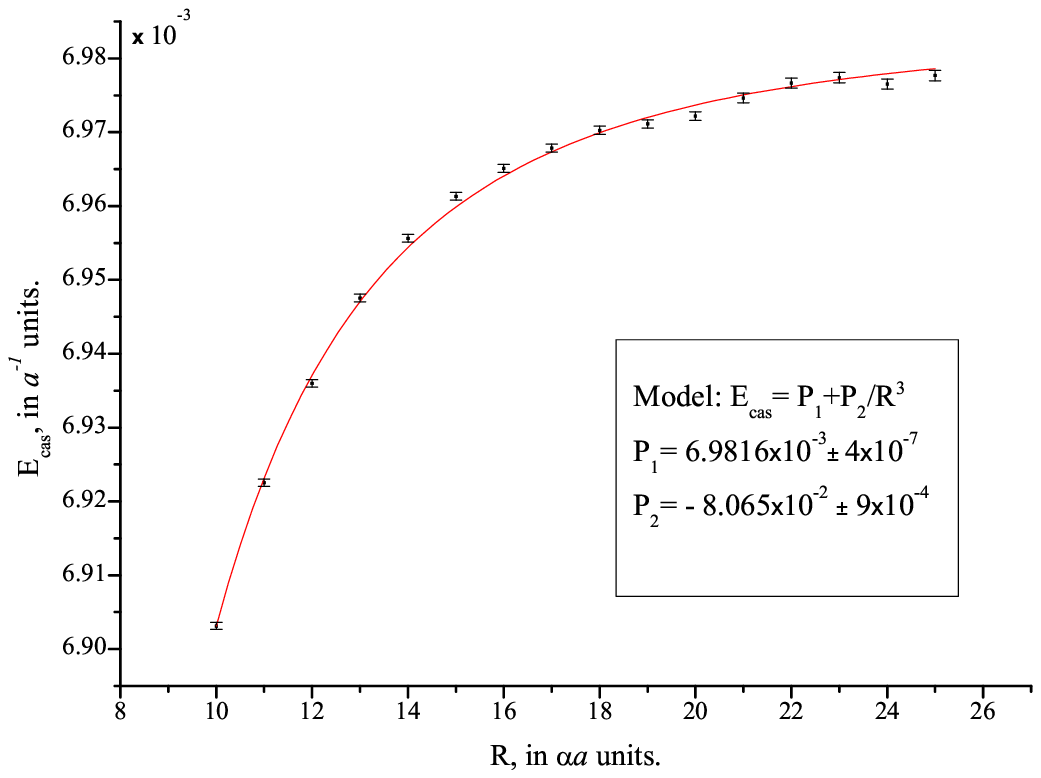} \end{center} \caption{$E_{cas}(R)$ dependence at $\beta=4.0$ and $\lambda=0.02$ on deformed lattice.
 Step in z-direction is $\alpha a$, in other directions it is $a$, $\alpha=0.2$.}
\end{figure}

The fitting procedure was realized again starting from distance
$R=2a=10 (\alpha a)$ (2 non-deformed steps is equal to 10 deformed
steps). As in the previous case, numerical points are very close
to the fitting function. There is one peculiarity which we want to
mention. The energy is expressed in the units $a^{-1}$, because we
didn't change the step in t-direction, but $R$ is expressed now in
the units $\alpha a$. So if the $E_{cas}(R)$ dependence is really
close to the $C/R^3$  function, $P_2$ coefficient in fitting
function (\ref{model}) on the deformed lattice should be divided
by $\alpha^3$ to obtain the $P_2$ coefficient on the non-deformed
lattice. And these numerical values of coefficients really satisfy
this condition.

The next step is the analysis of $P_2$ dependencies on the
parameter $\lambda$ and on the size of the lattice. The first one
is presented in the Fig. 10. In accordance with (\ref{assimpt}),
the calculated coefficient at $R^{-3}$ is proportional to
$\lambda^2$ for small $\lambda$. Unfortunately, numerical errors
increase rapidly for large $\lambda$ and in present calculations
we are limited to small values of $\lambda$.

\begin{figure}[t]
\begin{center}
 \epsfbox{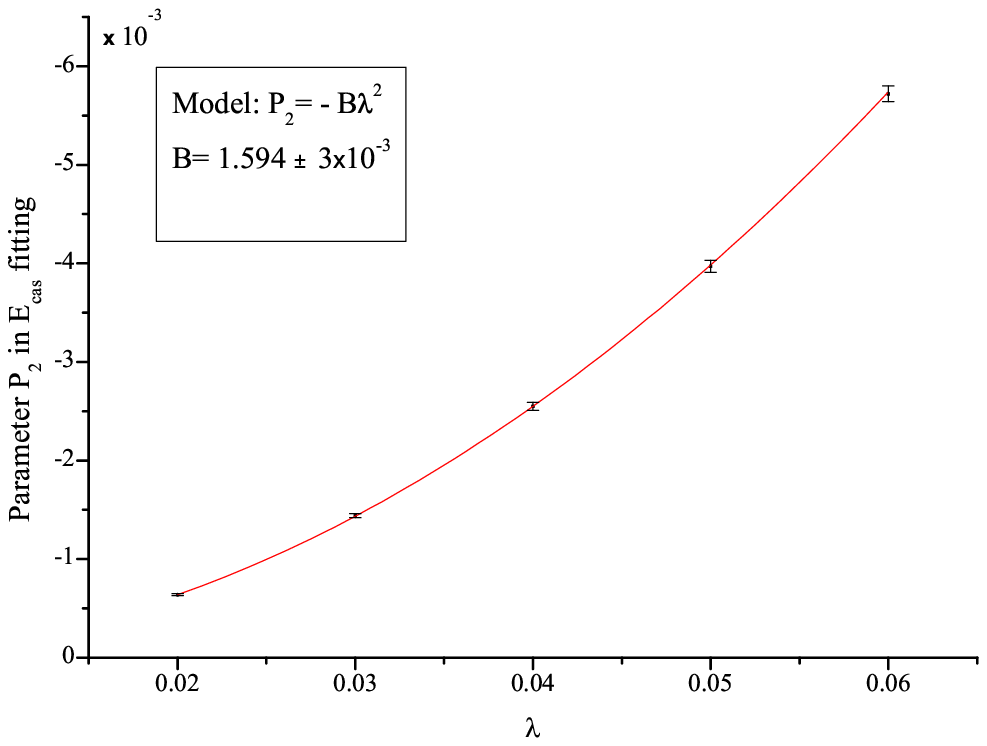} \end{center} \caption{$P_2(\lambda)$ dependence. Lattice is the same that was used in Fig. 1.}
\end{figure}

The dependence of $P_2$ on the lattice size is shown in Fig. 11.
$P_2$ is proportional to $N^2$, where $N$ is the size of the
lattice in x and y-direction. So the calculated Casimir energy of
two planes is really proportional to theirs area.
\begin{figure}[t]
\begin{center}
 \epsfbox{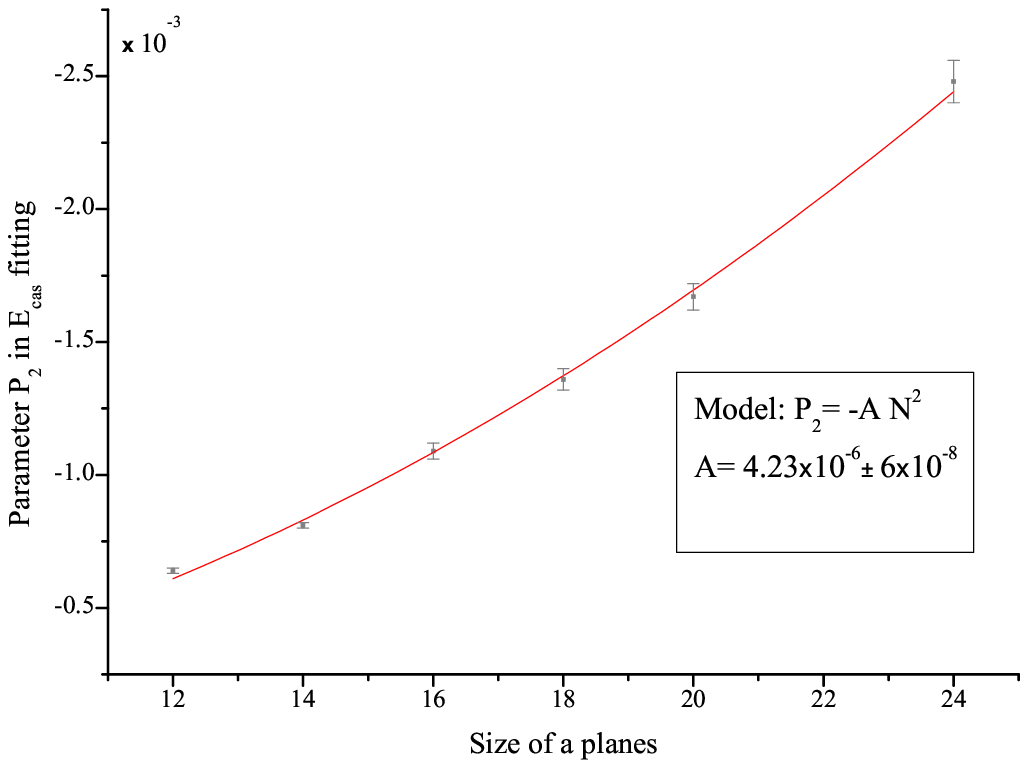} \end{center} \caption{$P_2(N)$ dependence. $N$ is the size of the lattice in x and y-directions.
 Lattice without deformations, $\beta=4.0$ and $\lambda=0.02$.}
\end{figure}

The last point is the consideration of the continuous limit of our
calculations. As we mentioned above nothing depends on the
parameter $\beta$, due to Casimir energy independence of electron
charge. So the only considerable stage in continuous limit is to
go to the "large lattice" limit where size of the lattice
$N\rightarrow \infty$. There is one feature in this continuous
limit procedure and we think that this moment should be
emphasized. The lattice step $a$ disappears from our
considerations because the final result of our calculation is the
dimensionless coefficient at $R^{-3}$ in expression for Casimir
energy per unit area. So we don't need to define the physical
volume of the lattice and the value of $a$ is not important for
us. Because of arguments discussed above the continuous limit for
$P_2$ coefficient is not sensitive to the limit $a \rightarrow 0$
and we have only the limit $N \rightarrow \infty$. $E_{cas}(R)$
dependence in Fig. 8. can be an illustration of this continuous
limit: the correct dependence $E_{cas}\sim R^{-3}$ reveals only at
lattice distances $R\geq2$.

\section*{Conclusions}

In this paper we have proposed the numerical method for the
Casimir energy calculation based on the lattice simulations of
QED. We have combined three ideas:
\begin{itemize}
\item The generation of the boundary conditions by means of the
additional Chern-Simons boundary action and new lattice
discretization of this additional action.

\item The lattice "Wilson bag" concept (the lattice presentation
of the closed 3-dimensions surface in Euclidian 4-dimensions
space).

\item The improvement of the numerical results by using deformed
lattice in space direction.
\end{itemize}

The combination of the first two ideas is in fact a lattice
definition of the respective quantum observable for the Casimir
energy and for the Casimir interaction between surfaces and last
one is a very important numerical tool to archive the necessary
accuracy.

We have tested our method in the simplest case of the Casimir
interaction between two plane surfaces and have achieved results
close to the analytical ones for this problem.

\section*{Acknowledgments}

We thank  Prof. V.V. Nesterenko, Dr. I.G. Pirozhenko, Dr. M.A.
Trusov and Dr. V.N. Marachevsky for interesting discussions about
the Casimir problem and Andrew Zayakin for essential remarks about
vortex dynamics of the Chern-Simons theory. The resources of MSU
supercomputer center were used in our calculations. The work is
partially supported by the Russian Federation President's Grant
195-2008-2.

\end{document}